\documentclass{iau}
\usepackage{graphicx}

\title[Using radio jets of PKS J0334-3900 to probe the intra-cluster medium in A3135] 
{Using radio jets of PKS J0334-3900 to probe the intra-cluster medium in A3135}

\author[L. Pratley et al.]   
{
L. Pratley $^1$,
M. Johnston-Hollitt$^1$\thanks{Presenter.},
S. Dehghan$^1$
 \and 
M. Sun$^2$
}

\affiliation{
$^1$School of Chemical \& Physical Sciences, Victoria University of Wellington, \\
PO Box 600, Wellington 6140, New Zealand \\ 
email: {\tt Melanie.Johnston-Hollitt@vuw.ac.nz} \\
[\affilskip]
$^2$Eureka Scientific, Inc., 2452 Delmer Street, Suite 100, Oakland, CA 94602, USA 
}

\pubyear{2014}
\volume{313}
\pagerange{xx}
\jname{Extragalactic jets from every angle}
\editors{F. Massaro, C.C. Cheung, E. Lopez, A. Siemiginowska, eds.}

\begin{document}

\maketitle

\begin{abstract}
We present a multi-wavelength study of the radio galaxy PKS J0334-3900, which resides at the centre of Abell 3135. Using Australia Telescope Compact Array (ATCA) observations at 1.4, 2.5, 4.6 \& 8.6 GHz, we performed a detailed analysis of PKS J0334-3900. The morphology and spectral indices give physical parameters that constrain the dynamical history of the galaxy, which we use to produce a simulation of PKS J0334-3900. This simulation shows that the morphology can be generated by a wind in the intracluster medium (ICM), orbital motion caused by a companion galaxy, and precession of the black hole (BH). 

Additionally, ATCA polarisation data was used to obtain rotation measure values along the line of sight to PKS J0334-3900. Using our simulation we are able to infer the distance between the jets along the line of sight to be 154 $\pm$ 16 kpc, which when combined with the difference in rotation measure between the jets provides a novel new way to estimate the average magnetic field within the cluster. A lower limit to the cluster magnetic field was calculated to be $0.09\pm 0.03$ $\mu$G. We have shown that different techniques can be applied to observations of jets in galaxies to infer information on cluster environments.

\end{abstract}

\firstsection 
\section{Introduction}
\begin{figure}
\begin{center}
\includegraphics[width=13cm]{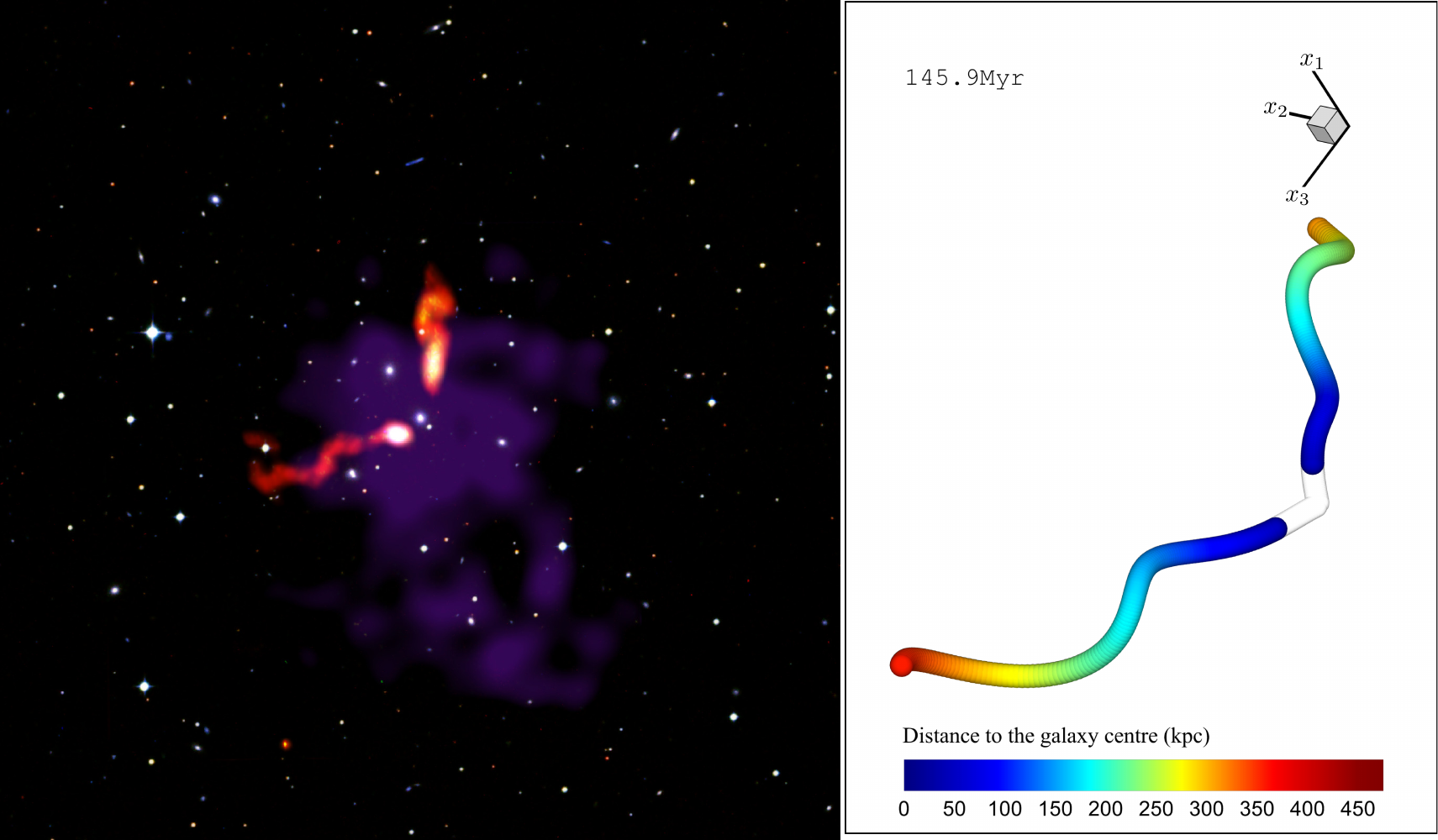}
 \caption{Left: Optical image of PKS J0334-3900 overlaid with the 1.4 GHz ATCA observation in orange and the Chandra 0.7 $-$ 2 keV observation in purple.  Right: Simulation of PKS J0332-3900, replicating the morphology of the jets due to a wind in the ICM, and the hook shapes generated by orbital motion with a nearby companion.}
   \label{fig1}
\end{center}
\end{figure}

PKS J0334-3900 is located at the centre of the galaxy cluster Abell 3135, as shown in Fig. \ref{fig1}. Abell 3135 is found at the northern edge of the Horologium-Reticulum supercluster, and has a redshift of $z=0.06228 \pm 0.00015$. PKS J0334-3900 is located at the same redshift with $z=0.062310 \pm 0.000097$. Chandra X-ray observations at 0.7 $-$ 2 keV show that the ICM in Abell 3135 is skewed to the southwest. It is likely that the system is in a late merging stage, and that the merger happened along the SW-NE.

Using flux measurements of PKS J0334-3900 between 80 MHz \& 8.6 GHz, the spectral index was found to be $\alpha=-0.72\pm 0.01$. Assuming a critical frequency of 178 MHz, the spectral index provides an estimated upper limit age of 146 Myr for the jet emission. The 1.4 GHz ATCA observations of PKS J0334-3900 shows the key features in morphology, the top and bottom jets are bent in the same direction, the jets have hooks towards the ends, and there are knots of intensity at the start of each jet. Additionally, the hook shapes are clearly seen in the magnetic component of the polarized emission. Furthermore, it is possible to see the magnetic component curve as it leaves the knot regions. At ${\rm RA_{J2000}}$ = 03:34:14, there is a break in emission in the bottom jet, suggesting the jet is changing direction along the line of slight. We hypothesized the jets are bent due to a wind in the ICM while the hooks are generated from orbital motion around a known companion galaxy combined with precession of the BH.

To test this hypothesis, we created a simulation of PKS J0334-3900 under the above conditions. The model of PKS J0334-3900 consists of symmetric and, to a first approximation, ballistic jets moving at constant speed (for details, see \cite[Pratley et al., 2013]{Pratley_etal13}). As shown in Fig. 1, we find that this model reproduces both the bending in the jets due to the wind in the ICM and the hooks due to the orbital motion around a companion galaxy and procession of the BH.

The ATCA observations at 1.4 GHz were also used to make a rotation measure (RM) image of PKS J0334-3900. The average rotation measure in the top and bottom jets are 1.16 and 7.44 rad m$^{-2}$, respectively. Using the simulation in Fig. 1, we estimated the jet separation to be $154\pm 16$ kpc along our line of sight. The Chandra observation provides an average electron density for the ICM of 1.09 cm$^{-3}$. We expect that the separation between the jets along our line of sight will cause the difference in average RM between the jets. In this case, a lower limit of the average cluster magnetic field is calculated as $0.09 \pm 0.03$ $\mu$G. (See \cite[Pratley et al. 2013]{Pratley_etal13} for full details.)

\section{Conclusion}

This work shows that the morphology of the jets in the bent-tailed galaxy PKS J0334-3900 can be explained by a wind in the ICM, precessional and orbital motion and that the combination of the data and simulation can be used to estimate the strength of the cluster magnetic field. In particular, this work highlights the way in which multi-wavelength observations of radio jets are powerful tools for probing their environments. Next generation radio telescopes will unveil bent-tailed sources in great numbers paving the way for environmental studies of this nature on large scales.

\end{document}